\documentclass{aa}  

\usepackage{graphicx}
\usepackage{txfonts}
\usepackage{xcolor}

\newcommand{\lppr}{\stackrel{<}{\scriptstyle \sim}}

\newcommand{\msun}{${\rm M}_{\sun}$}

\usepackage{hyperref}
\begin{document}

   \title{White dwarf spectral type--temperature distribution from {\it Gaia}-DR3 and the Virtual Observatory}

   \subtitle{}

\author{S. Torres\inst{1,2}\thanks{Email;
    santiago.torres@upc.edu}\and
        P. Cruz\inst{3}\and
        R. Murillo-Ojeda\inst{3}\and
        F. M. Jim\'enez-Esteban\inst{3}\and
        A. Rebassa--Mansergas\inst{1,2}\and 
        E. Solano\inst{3}\and
        M. E. Camisassa\inst{1}\and
        R. Raddi\inst{1}\and
        J. Doliguez Le Lourec\inst{4}
        }
        
\institute{Departament de F\'\i sica, 
           Universitat Polit\`ecnica de Catalunya, 
           c/Esteve Terrades 5, 
           08860 Castelldefels, 
           Spain
           \and
           Institute for Space Studies of Catalonia, 
           c/Gran Capit\`a 2--4, 
           Edif. Nexus 104, 
           08034 Barcelona, 
           Spain
           \and
 Centro de Astrobiolog\'{\i}a (CAB), CSIC-INTA, Camino Bajo del Castillo s/n, campus ESAC, 28692, Villanueva de la Ca\~nada, Madrid, Spain
          \and
 \'Ecole d'ing\'enieurs (EPF), 
  55 Av. du Pr\'esident Wilson, 
  94230 Cachan,  
  France}
           \date{\today}

 
  \abstract
   {The characterization of white dwarf atmospheres is crucial for accurately deriving stellar parameters such as effective temperature, mass, and age. However, the inclusion of physical processes like convective mixing and convective dilution in current white dwarf atmospheric models predicts a spectral evolution of these objects. To constrain these models, accurate observational data and analysis are necessary.}
   {To classify the population of white dwarfs up to 500 pc into hydrogen-rich or hydrogen-deficient atmospheres based on {\it Gaia} spectra and to derive an accurate spectral type-temperature distribution, i.e., the ratio between the number of non-DAs to the total number of white dwarfs as a function of the effective temperature for the largest observed unbiased sample of these objects.}
   {We took advantage of the recent {\it Gaia} low-resolution spectra available for 76\,657 white dwarfs up to 500 pc. We calculated synthetic J-PAS narrow-band photometry and fitted the spectral energy distribution of each object with up-to-date models for hydrogen-rich and helium-rich white dwarf atmospheres. We estimated the probability for a white dwarf to have a hydrogen-rich atmosphere and validated the results using the Montreal White Dwarf Database.  Finally, precise effective temperature values were derived for each object using La Plata evolutionary models.}
   {We have successfully classified a total of 65\,310 white dwarfs (57\,155 newly classified objects) into DAs and non-DAs with an accuracy of 94\%. An unbiased subsample of nearly 34\,000 objects was built, from which we computed a precise spectral distribution spanning an effective temperature range from 5\,500 to 40\,000 K, while accounting for potential selection effects.}
   {Some characteristic features of the spectral evolution, such as the deficit of helium-rich stars at $T_{\rm eff} \approx 35\,000-40\,000\,$K and in the range $22\,000 \lesssim T_{\rm eff} \lesssim 25\,000\,$K, as well as a gradual increase from $18\,000\,$K to $T_{\rm eff} \approx 7\,000\,$K, where the non-DA stars percentage reaches its maximum of 41\%, followed by a decrease for cooler temperatures, are statistically significant. These findings will provide precise constraints for the proposed models of spectral evolution.}

\titlerunning{White dwarf spectral distribution}
\authorrunning{Torres et al.}

\maketitle

%

\section{Introduction}
\label{s:intro}

The {\it Gaia} mission has revealed important features of the white dwarf population with unprecedented precision. That is the case, for instance, of the existence of two well-defined branches in the color-magnitude diagram (roughly between $0.0\lesssim G_{\rm BP}-G_{\rm RP}\lesssim0.5$) referred to as the A and B branches, which correspond to the majority of white dwarfs with hydrogen-rich and helium-rich atmospheres \citep[see][]{Gaia2018}. The most plausible explanation to reproduce the B branch invokes the presence of small amount of hydrogen or carbon into helium-dominated atmospheres \citep{Bergeron2019,Camisassa2023,Blouin2023}. These models are based on well-studied physical processes that can alter the composition of the outer layers, such as convective mixing and convective dilution, among others \citep[e.g.][and references therein]{Rolland2018}. The specific characteristics of each model, such as the hydrogen content, the depth of the convective zone as a function of the effective temperature, or even the possibility of accreting material from surrounding asteroids, give rise to different channels of formation and evolution of white dwarf spectral types\footnote{White dwarfs are classified as DAs or non-DAs based on the presence or absence of hydrogen lines in their spectra, respectively. This last group is formed by those who exhibits helium lines (DB), metal lines (DZ), carbon lines (DQ), or no lines at all (DC), among others \citep{Sion83}.} \citep[e.g.][]{Rolland2018,Ourique2018,Cunningham2019,Cunningham2020,Bedard2020}. Therefore, accurate observational data is required to constrain these models. \\

\begin{figure*}[t]

      {\includegraphics[width=0.95\columnwidth, clip=true,trim=-100 0 0 0]{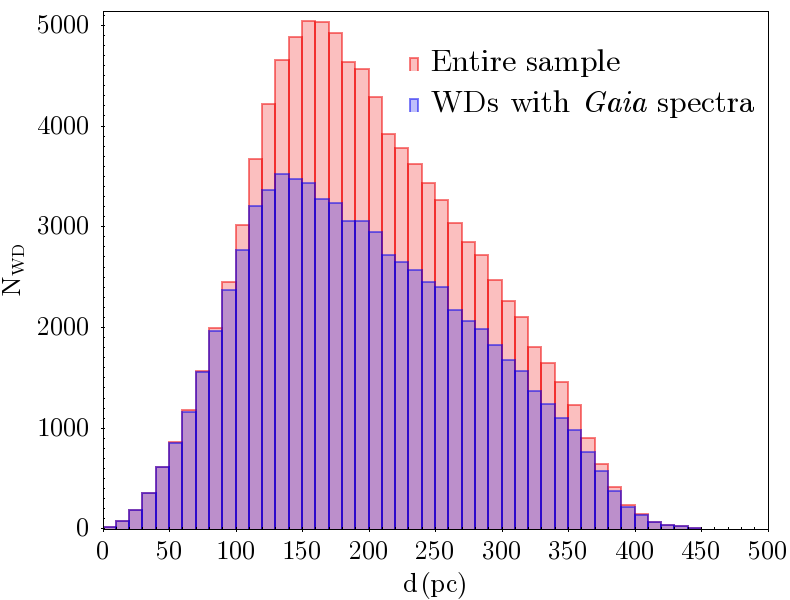}}
      {\includegraphics[width=0.95\columnwidth, clip=true,trim=-100 0 0 0]{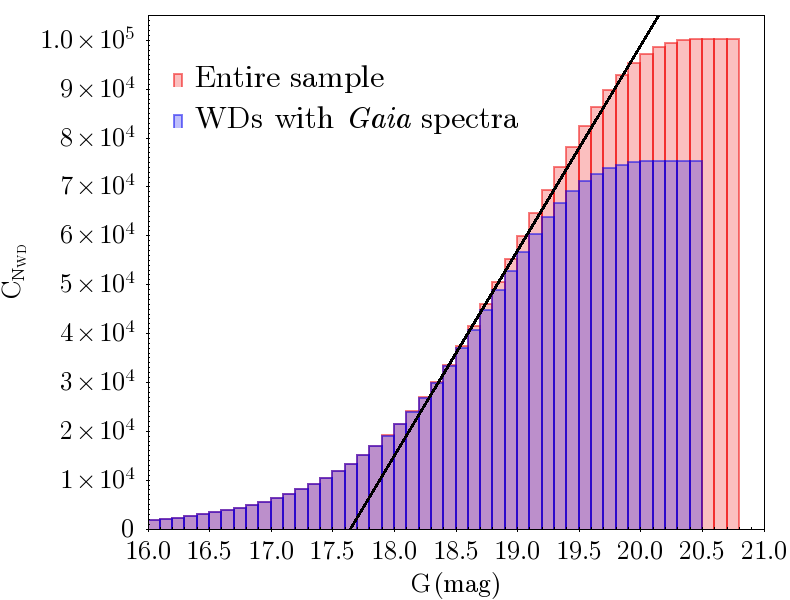}}

\caption{Distance distribution (left panel) and apparent $G$ magnitude cumulative distribution (right panel) for the entire sample of white dwarfs that fulfil our selection criteria (red histogram) and for the subsamble of objects that have {\it Gaia} spectra (blue histogram). A constant cumulative slope is shown (black line) as indicative of the completeness of the sample. }
\label{f:distri}
\end{figure*}

The proper explanation for the formation of the {\it Gaia} A and B branches extends beyond the effective temperature range of these branches, encompassing a broader issue.  A crucial observational factor in analyzing the spectral evolution of white dwarfs is the ratio of non-DA to DA stars as a function of effective temperature\footnote{Throughout the paper, we use the term 'spectral type-temperature distribution' or simply 'spectral distribution' to refer to the ratio of non-DA to the total number of objects as a function of the effective temperature. We prefer it to the term 'spectral evolution' as we consider this case specifically refers to the physical processes that result in a change in spectral type.}. Extensive efforts have been made since the pioneering work of \citet{Sion1984} to obtain statistically significant spectral distributions \citep[e.g.][]{Tremblay2008}. However, the advent of large spectroscopic and photometric surveys such as the Sloan Digital Sky Survey (SDSS; \citealt{Sloan2000}), {\it Galaxy Evolution Explorer} ({{\it GALEX}; \citealt{Morrissey2007}), and {\it Gaia} \citep{Prusti2016} has significantly increased both the quantity and quality of available data \citep[e.g.][]{Ourique2018,Genest2019,Blouin2019,Cunningham2020,McCleery2020,Lopez-Sanjuan2022}. Even though, complete spectroscopic samples have been limited to a distance of up to 40 pc, or in other cases, magnitude-selection effects introduce significant biases in the final distribution.

Nevertheless, we can leverage the exceptional quality of astrometric and photometric data provided by the {\it Gaia} mission. The third data release (DR3) of {\it Gaia} includes low-resolution spectra for nearly 100\,000 white dwarfs \citep{Gaia2022}, making it the largest sample of white dwarfs available for analysis. In our recent study \citep{JE+2023}, we classified 8\,150 white dwarfs within a nearly volume-complete 100\,pc sample into DA or non-DA categories. The achieved accuracy of 90\% was remarkable and allowed us to derive a detailed spectral distribution within the range of temperatures from $5\,500\,$K up to $23\,000\,$K.

In this paper, we extend our previous analysis to a distance of 500 pc, significantly increasing the expected number of white dwarfs, particularly for hotter effective temperatures. Our goal is to derive for the first time the spectral distribution in the entire range of temperatures between $5\,500\,$K up to $40\,000\,$K, where spectral evolution is significant.

The paper is structured as follows: in Section 2, we describe the selection procedure used to obtain our white dwarf sample from {\it Gaia} data. We provide a summary of the main steps of our classification methodology in Section 3. Once our sample is classified and validated, Section 4 focuses on analyzing selection effects and addressing the completeness correction of the sample. In Section 5, we present our spectral distribution, discuss our findings, and compare them to previous works. We summarize our key results and draw our main conclusions in Section 6.

\section{The {\it Gaia}-DR3 white dwarf sample}
\label{s:sample}

We selected our objects from  {\it Gaia}-DR3 catalogue\footnote{\url{http://gea.esac.esa.int/archive/}} following the criteria used in \citet{JE+2023} but extended up to 500 pc:
\begin{itemize}
  \item $\omega-3\sigma_{\omega}\ge 2$\,mas and $\omega/\sigma_{\omega}\ge10$
\item $F_{\rm BP}/\sigma_{F_{\rm BP}}\ge10$ and $F_{\rm RP}/\sigma_{F_{\rm RP}}\ge10$
  \item RUWE<1.4; where RUWE stands for Renormalised Unit Weight Error preventing against poor astrometric solutions \citep{Lindegren20}.
  \item $|C^{*}|<3\sigma_{C^{*}}$; where $|C^{*}|$ is an estimate of the BP and RP flux excess factor and $\sigma_{C^{*}}$ its scatter following the prescription by \cite{Riello20}.
 \end{itemize}

Selected objects were corrected from extinction following the 3D interstellar Galactic extinction maps from \citet{Lallement2022}\footnote{\url{https://stilism.obspm.fr/}}. 
In principle, we selected only those objects falling below the 0.45\,\msun\  cooling track on the Hertzsprung-Russell (HR) diagram. Additionally, as the atmospheric models we used (see Section \ref{s:class}) provide a reliable estimate of the effective temperature in the range from 5\,500\,K up to 40\,000\,K, we chose those objects with unreddened color between $-0.5<{\rm BP-RP}<0.86$. A total number of 100\,173 objects were selected, from which 76\,657 have {\it Gaia} low-resolution spectra available.

\begin{figure*}[t!]
      {\includegraphics[width=0.95\columnwidth, clip=true,trim=-100 0 0 0]{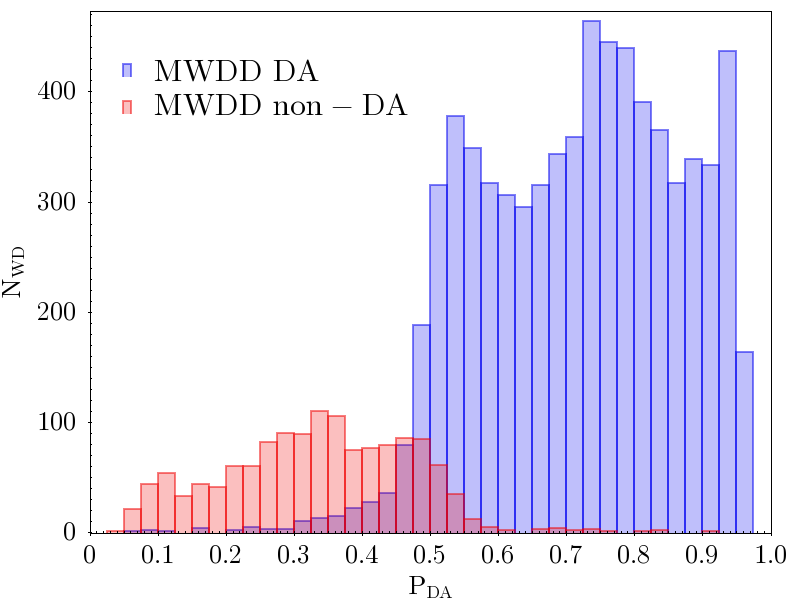}}
      {\includegraphics[width=0.95\columnwidth, clip=true,trim=-70 10 10 -10]{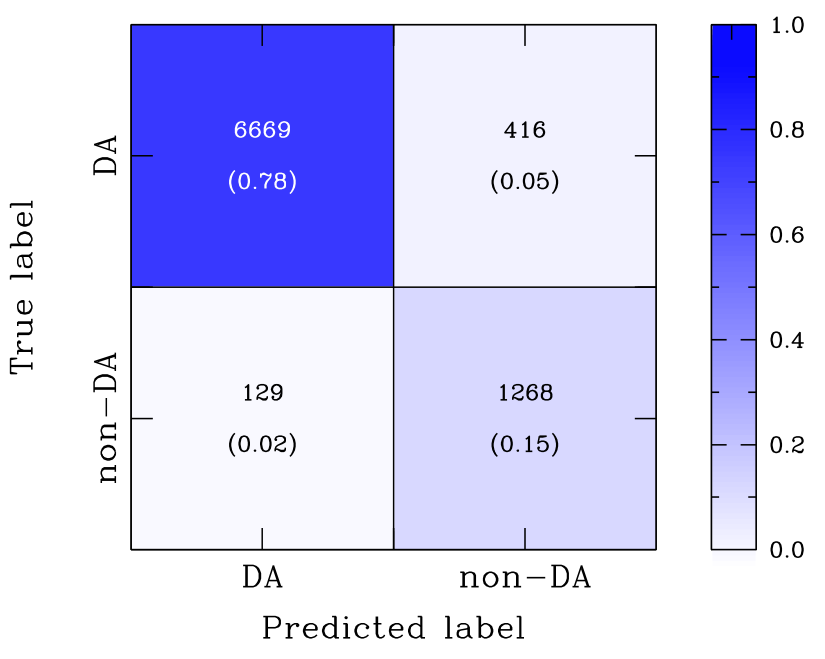}}

\caption{{\sl Top panel:} probability distribution of being DA for the white dwarf labelled as DA or non-DA  (blue and red histograms, respectively) in the MWDD. {\sl Bottom panel:} confusion matrix of our estimator of being DA. Displayed values represent the total number of objects, while in brackets
the percentages with respect to the total population.}
\label{f:PDA_MWDD}
\end{figure*}

Figure \ref{f:distri} displays the distance distribution (left panel) and cumulative distribution of apparent magnitude $G$ (right panel) for our entire sample (red histogram) and white dwarfs with {\it Gaia} spectra (blue histogram). The use of inverse parallax as a distance estimator, combined with a parallax error threshold of less than 10\%, introduces negligible discrepancies (less than $\approx 4\%$) compared to other distance estimators \citep{Bailer-Jones2021}. Most of our selected white dwarfs (73\%) are within 250 pc, with a long tail extending up to 500 pc. The cumulative $G$ magnitude distribution reveals a deficit of objects starting at $G\sim 20$\,mag for the entire sample and around $G\sim 19.5$\,mag for white dwarfs with {\it Gaia} spectra. Although the nominal limiting {\it Gaia} magnitude is $\sim 21.0$, we adopted a conservative value of $G_{\rm lim}=19.5$\,mag for our completeness analysis (see Section \ref{s:comp}).

\section{White dwarf spectral classification}
\label{s:class}

For those sources of our selected sample with available {\it Gaia} spectra, we followed the same procedure as described in \citet{JE+2023}. A brief description of the methodology  for classifying white dwarfs into DA and non-DA types used in that work is provided as follows.

 First, for each white dwarf of our sample with available {\it Gaia} spectrum we determined, by means of the Python package {\it GaiaXPy}\footnote{\url{https://www.cosmos.esa.int/web/gaia/gaiaxpy}} and taking into account all the coefficients of the {\it Gaia} spectrum, the synthetic Javalambre-Physics of the Accelerating Universe Astrophysical Survey (J-PAS; \citealt{Benitez14}) filter system \citep{Marin-Franch12} photometry. 
We focused on those filters covering the range from 4000 to 9590.54\,\AA, and discarding those filters with effective wavelength shorter than 4000\,\AA.  Second, for each object we built a spectral energy distribution (SED) using the derived J-PAS photometry.  Although most of the SEDs have 56 photometric points, in some noisy spectra the number of points is lower, due to  our threshold in the photometric error of 10\% to each individual photometric measurement obtained with {\it GaiaXPy}.

We analyzed $67\,340$ new white dwarfs spectra, not previously studied in our 100 pc sample \citep{JE+2023}. For those objects with more than 4 photometric points ($57\,155$), their SEDs were fitted using either pure hydrogen white dwarf atmospheric models (DA) or models with helium and a small trace of hydrogen (non-DA, log N(H)/N(He) = -6). Both DA and non-DA models covered the temperature range of interest for this study ($5\,500\,$ to $40\,000\,$K) and surface gravities from 7 to 9 dex (see Section 3.1 in \citealt{JE+2023} for detailed model information). The fitting process was performed using the Virtual Observatory Spectral energy distribution Analyzer\footnote{\url{http://svo2.cab.inta-csic.es/theory/vosa}} (VOSA; \citealt{Bayo08}), a powerful tool developed by the Spanish Virtual Observatory. Among the new $57\,155$ analyzed objects, only 3 had Vgf$_b$\footnote{Vgf$_b$: Modified reduced $\chi^{2}$ calculated by forcing $\sigma(F_{\rm obs})$ to be larger than $0.1\times F_{\rm obs}$, where $\sigma(F_{\rm obs})$ is the error in the observed flux ($F_{\rm obs}$). Vgf$_b$ smaller than 10--15 is often perceived as a good fit.} greater than 15 and were disregarded. For each object, two reduced chi-squared values ($\chi^2_{\rm DA}$ and $\chi^2_{\rm non-DA}$) were obtained. Finally, the estimator of the probability of being a DA white dwarf ($P_{\rm DA}$) was defined as
\begin{equation}
\label{e:probDA}
P_{\rm DA}=\frac{1}{2}\left(\frac{\chi^2_{\rm non-DA}-\chi^2_{\rm DA}}{\chi^2_{\rm non-DA}+\chi^2_{\rm DA}}+1 \right),
\end{equation}
where we classified an object as DA if $P_{\rm DA}\ge 0.5$, otherwise as a non-DA.

We validated our classification procedure by means of the spectroscopically labelled white dwarf sample of the Montreal White Dwarf Database\footnote{\url{https://www.montrealwhitedwarfdatabase.org/}} (MWDD; \citealt{Dufour2017}). A total of 8\,482 objects from the MWDD were used in our validating test (including those at distances closer than 100\,pc). We adopted as DA class all MWDD objects whose primary spectral type is DA regardless  of secondary types. The rest of the objects were considered as non-DA. In the left panel of figure \ref{f:PDA_MWDD} we show the distribution of the probability of being a DA, $P_{\rm DA}$, for white dwarfs labelled as DA (blue histogram) or non-DA (red histogram) in the MWDD.  The distribution confirms that the adopted threshold at $P_{\rm DA}=0.5$ effectively separates both populations and that the percentage of missclassified object is reasonable small, $\lesssim 7\%$.  In the right panel of Fig. \ref{f:PDA_MWDD} we show the confusion matrix, where rows  represent the number of already labelled spectral objects, while columns are the prediction of our classification (in parentheses the percentages with respect to the total population). The results of the confusion matrix  reveal that the performance of our spectral type estimator is excellent, as it is corroborated by the derived metrics\footnote{For a definition of these parameters, please refer for instance to Appendix A from \cite{Echeverry2022}}: accuracy of 0.94, F1-score of 0.96, recall of 0.94 and precision of 0.98.

\begin{figure}[h!]
      {\includegraphics[width=0.93\columnwidth, clip=true,trim=-50 0 0 0]{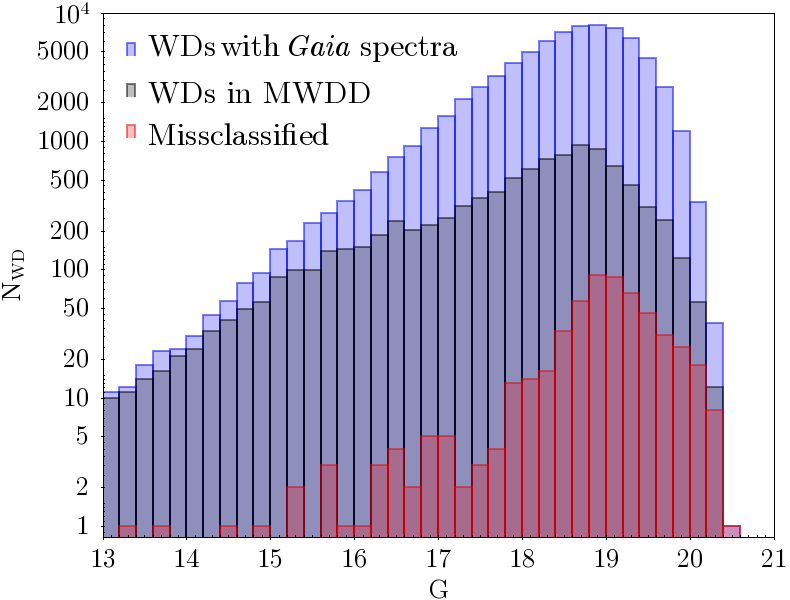}}

\caption{Distribution of $G$ apparent magnitude for the entire sample of white dwarfs with {\it Gaia} spectra (blue histogram), those with spectral classification in MWDD (gray histogram) and those of the previous sample missclassfied by our method (red histogram).}
\label{f:Gapa_dis}
\end{figure}

A last check was performed before applying our classification method to the observed {\it Gaia} white dwarf sample. In Figure \ref{f:Gapa_dis} we depicted the distribution of the $G$ apparent magnitude for the entire sample of white dwarfs with {\it Gaia} spectra (blue histogram) and  those with spectral classification in MWDD (gray histogram). We verified that the MWDD sample covers  the full range of magnitudes and closely resemble the observed {\it Gaia} sample distribution.  These facts guarantee the proper use of the MWDD sample for testing our classification method. Moreover, in Fig. \ref{f:Gapa_dis}, we present the magnitude distribution of the white dwarfs misclassified by our method (red histogram).  As expected, the fainter the magnitude the larger the fraction of missclassified objects. Considering the percentage of error as a function of magnitude and extrapolating it to the {\it Gaia} sample, we estimated that the error in the final classification should not exceed 10\%.

Once we have validated  the reliability of our classification method, we applied it to the sample of white dwarfs with {\it Gaia} spectra within 500\,pc. A total of 65\,310 white dwarfs (including those previously classified in \citealt{JE+2023}) have been classified into the spectral types DA (50\,189; 77\%) and non-DA (15\,121; 23\%) with an accuracy of 0.94.
The catalogue with the spectral classification is available online as supplementary material hosted by the journal and at \emph{The SVO archive of Gaia white dwarfs}\footnote{http://svocats.cab.inta-csic.es/wdw/index.php} at the Spanish Virtual Observatory portal\footnote{https://svo.cab.inta-csic.es/docs/index.php?pagename=Archives}.

\begin{figure*}[ht!]

      {\includegraphics[width=0.95\columnwidth, clip=true,trim=-80 0 0 0]{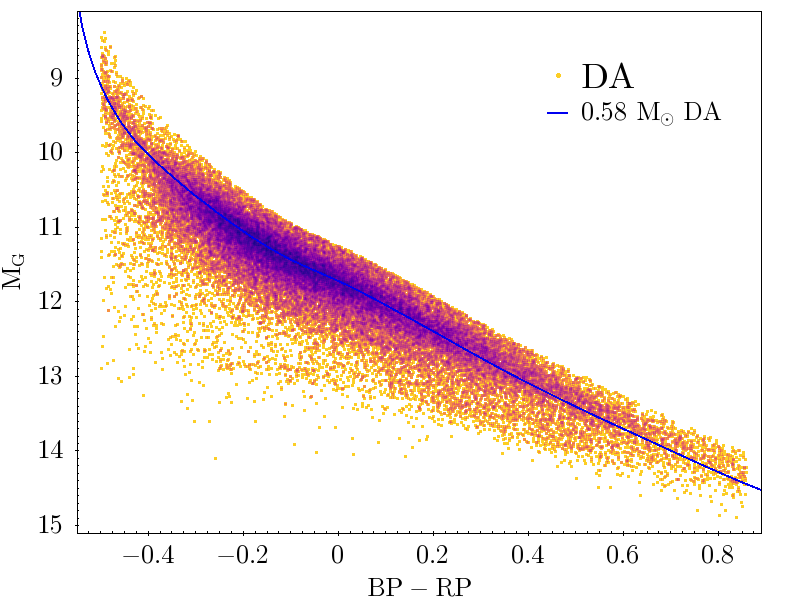}}
      {\includegraphics[width=0.95\columnwidth, clip=true,trim=-80 0 0 0]{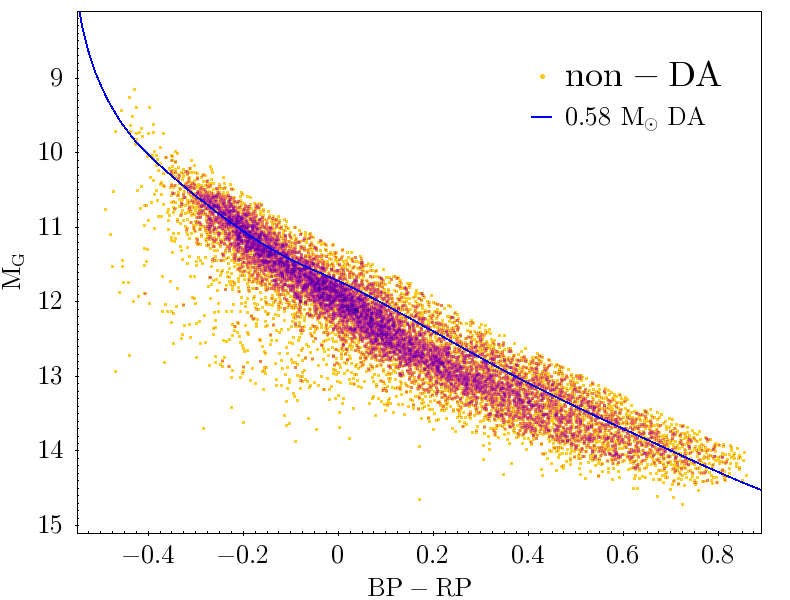}}

\caption{{\it Gaia} HR diagram for the population of white dwarfs classified by our probability estimator as DA (left panel) and non-DA (right panel). As a visual reference, we plotted the cooling sequence of a 0.58\,\msun\ DA white dwarf according to La Plata models.}
\label{f:HRDAnDA}
\end{figure*}

 In Figure \ref{f:HRDAnDA}, we showed the HR diagram of the corresponding DA and non-DA populations. For visual reference we also showed the cooling sequence of a 0.58\,\msun\ DA white dwarf \citep{Camisassa2016}. The distribution of DA and non-DA white dwarfs clearly follows different tracks, with the latter group being, on average, less luminous for a certain color (in particular for ${\rm BP-RP}>0$) than the first one. It is worth mentioning that our classification into DA and non-DA groups is less model dependent than H- versus He-rich classification. However, such a classification is still needed for a correct astrophysical interpretation. Thus, higher resolution spectroscopy is required to flag misclassified objects, identify He-rich DAs, magnetic DAH with distorted Balmer jumps, as well as DZ and DQ with unusual colors, among other cases.

\section{Completeness correction}
\label{s:comp}

\begin{figure*}[t]

      {\includegraphics[width=0.95\columnwidth, clip=true,trim=-80 0 0 0]{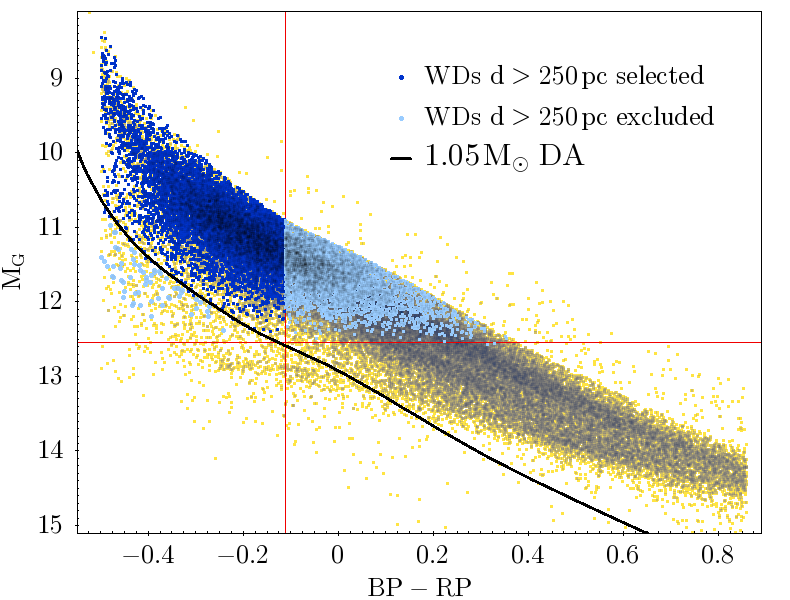}}
      {\includegraphics[width=0.95\columnwidth, clip=true,trim=-80 0 0 0]{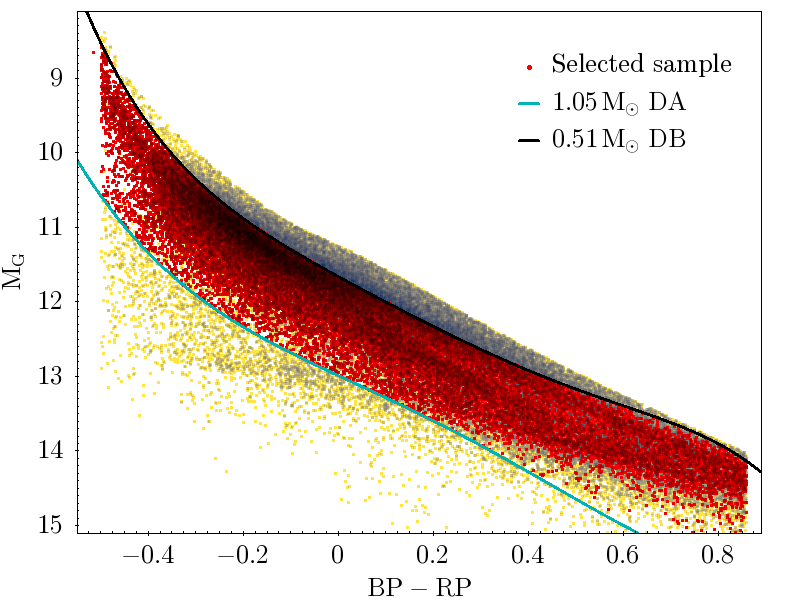}}

\caption{{\sl Left panel:} {\it Gaia} color-magnitude diagram of our population of DA and non-DA white dwarfs (yellow dots). Highlighted in blue are those objects at distances $d>250\,$pc. Assuming a limiting magnitude of $G_{\rm lim}=19.5$, only those  brighter  than $M_G<12.5$ (horizontal red line) are observable.  The cooling track for a 1.05\,\msun white dwarf (black line) is adopted as our lower selection function limit. For a given distance, only objects (marked in dark blue) at the left of the corresponding color value (vertical red line) contribute to the final spectral distribution. {\sl Right panel:} highlighted in red are the white dwarfs selected for building our spectral distribution. Objects above or below the cooling track for a 0.51\,\msun\ helium-rich and a 1.05\,\msun\ hydrogen-rich white dwarf, respectively, were discarded.}
\label{f:HR_comple}
\end{figure*}

We analyzed the different selection effects and how they can be corrected or at least mitigated. First of all, as our sample is initially built as a magnitude-selected sample, objects fainter than a certain magnitude limit would be absent from our sample. A standard $1/\mathcal{V}_{\rm max}$ method \citep{Schmidt1968} will provide an unbiased estimate of the space density. However, it requires extra conditions of completeness and homogeneity to be fulfilled \citep[see][and references therein]{Geijo2006}. These conditions are not guaranteed in our sample, as the requirement to have a {\it Gaia} spectrum adds a new selection effect and, mainly because DA and non-DA populations have, as previously stated, a different distribution in the color-magnitude diagram (see Fig.\ref{f:HRDAnDA}). 

In order to avoid this bias that would distort the spectral distribution, we developed a strategy in which we consider objects contributing to the spectral distribution only if they are brighter than a certain magnitude and hotter than a certain temperature (color). We adopted the cooling sequence for a $1.05\,$\msun\  white dwarf as our faint limiting region. For a given distance of the white dwarf, the {\it Gaia} limiting magnitude we adopted $G_{\rm lim}=19.5$ (see Section \ref{s:sample}) will fix the absolute observable magnitude limit. The corresponding color at this magnitude for the 1.05\,\msun\  white dwarf cooling sequence delimits the possible contribution to the spectral distribution. Only objects with a bluer color than this value will contribute, while redder objects will be disregarded. In the left panel of Figure \ref{f:HR_comple}, we present an example of this strategy. White dwarfs beyond a distance of 250\,pc are highlighted in the {\it Gaia} HR diagram. Only those objects (marked in blue) above the horizontal line are observable, and those to the left of the vertical line and above the 1.05\,\msun\  track are included in the final sample to construct our spectral distribution. Thus, we prevent non-DA objects (which, on average, are fainter than DAs for a given color; see Fig. \ref{f:HRDAnDA}) from being underestimated in the spectral distribution. It should be noted that this procedure automatically eliminates any massive white dwarf from our sample. However, their contribution is estimated to be less than $\approx 3$\% of the entire population \citep[e.g.][]{Kilic2020,JE+2023}.

\begin{figure}[t]
      {\includegraphics[width=0.97\columnwidth, clip=true,trim=0 10 20 10]{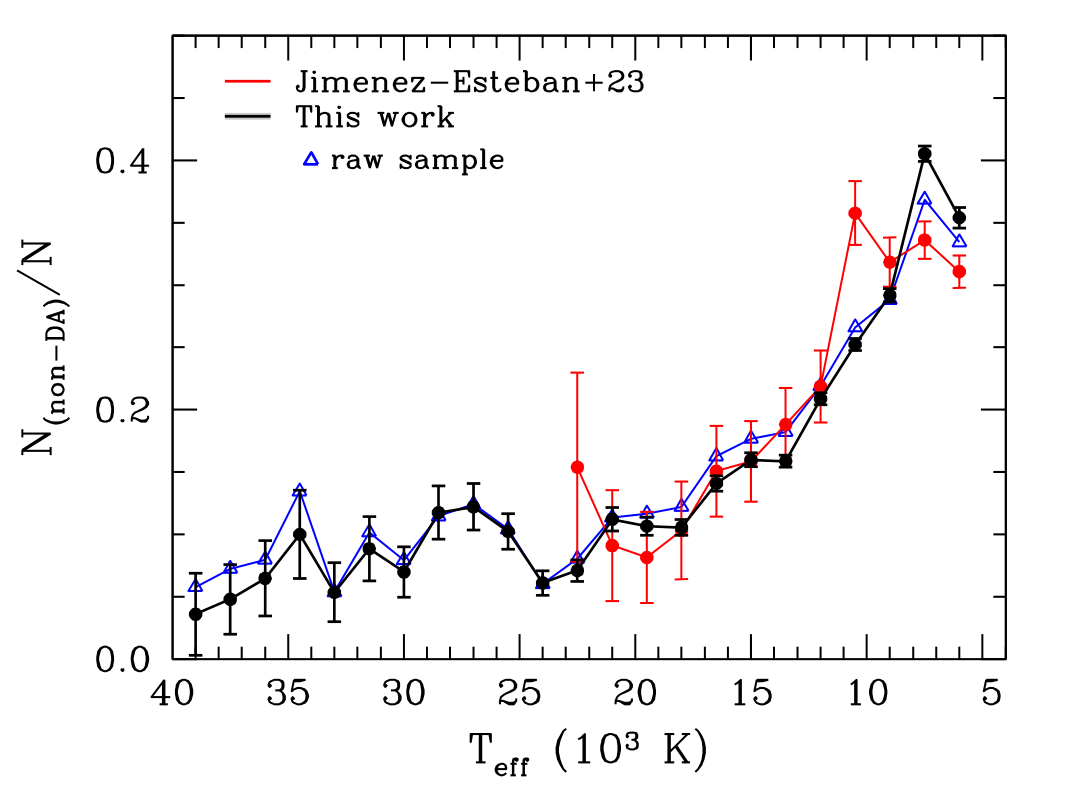}}

\caption{Spectral distribution presented in this work (black lines and solid circles) and when no selection criteria and correction function is applied at all (blue open triangles). Also shown the one obtained for the 100 pc sample \cite[][red circles]{JE+2023}.}
\label{f:raw}
\end{figure}

A second important source of incompleteness comes from the fact that not all white dwarf sources have an available {\it Gaia} spectrum. Moreover, even those sources that have it may not have a good VOSA determination of the probability of being DA or non-DA. It is expected that the number of sources without a determination of this probability increases for fainter and distant objects. We take into account this fact by introducing a weight function that depends on the distance, $d$, of the object and its specific location within the HR diagram, $w(G_{\rm BP}-G_{\rm RP},\, M_G,\,d)$. For a given source with parameters $(G_{\rm BP}-G_{\rm RP},\, M_G,\,d)_0$ we computed the number of sources, $n_{\rm sources}$, inside a volume $\mathcal{V}=\Delta(G_{\rm BP}-G_{\rm RP})\times\Delta M_G\times\Delta d$ centered at the previous value and with $\Delta(G_{\rm BP}-G_{\rm RP})=0.1$, $\Delta M_G=0.1$ and 
$\Delta d=50\,$pc. Besides the number of sources inside $\mathcal{V}$, we also computed the number of objects with available {\it Gaia} spectra, $n_{Gaia-sp}$, and the number of those who have a VOSA estimation of the probability of being DA, $n_{VOSA-PDA}$. Assuming that the completeness weight function should be inversely proportional to the probability of an object of belonging to the final sample, a straightforward application of the Bayes' theorem for conditional probability leads to: 

\begin{equation}
w(G_{\rm BP}-G_{\rm RP},\, M_G,\,d)=\left(\frac{n_{Gaia-sp}}{n_{sources}}\times \frac{n_{VOSA-PDA}}{n_{Gaia-sp}}\right)^{-1}=\frac{n_{sources}}{n_{VOSA-PDA}}.
\end{equation}

Finally, in addition to the selection function previously described (that is, we selected objects hotter than a given color determined by the $1.05\,$\msun\ evolutionary track for a given distance), we considered objects located in the HR diagram below the cooling track for a $0.51\,$\msun\ helium-rich white dwarf. 
This way, we avoid unresolved binary white dwarf systems and the contribution of white dwarfs evolved from binary evolution \citep{JE+2023}. Furthermore, the existence of  low-mass white dwarfs with helium-rich atmospheres has not been proven \citep[see, for instance,][]{Genest2019,Battich2020}. Thus, we adopted the $0.51\,$\msun low-mass limit as a conservative criterion. For each object of our final sample we determined the effective temperature by interpolating the {\it Gaia} photometry in the La Plata models. Those objects classified as DA were interpolated in the models of \citet{Camisassa2016}, while for those labelled as non-DA,  we used the hydrogen-deficient cooling models of \citet{Camisassa2017}, in both cases for carbon-oxygen-core white dwarfs \citep[see][for details]{JE+2023}. Atmospheric models where those used in the SED analysis \citep[see Section 3.1 from][]{JE+2023}, i.e., Koester's models with pure hydrogen composition for DAs and helium with a small trace of hydrogen (log N(H)/N(He) = -6) for non-DAs \citep{Koester10}. It is worth noting here that recent studies have emphasized the importance of considering the carbon content in non-DA atmospheres \citep{Camisassa2023,Blouin2023}. Assuming the maximum non-observable carbon enrichment prescription from \cite{Camisassa2023}, i.e. carbon sequence -1 dex, the difference with respect to a pure helium model is  $\sim10$\% at 12\,000\,K, and approximately 5\% at $6\,000\,$K. The differences are not larger than $1\,500\,$K, which corresponds to the bin width of our spectral distribution. Thus, no major effects are expected in this regard.

Our final sample to estimate the spectral fraction, consisting of 33\,997 white dwarfs, is shown in Figure \ref{f:HR_comple}, with 25\,984 (76.4\%) classified as DAs and 8\,013 (23.6\%) as non-DAs.

\section{The spectral type-temperature distribution}

The spectral type-temperature distribution, $f$, is defined as the ratio of weighted non-DA white dwarfs, to the total number of weighted objects, $N_w$, per effective temperature interval.  We adopted that the contribution of each object to its temperature bin depends on its weight function, $w$, and the probability of classification, $P_i$ (that is $P_{\rm DA,i}$ for DAs, $1-P_{\rm DA,i}$ for non-DAs). Hence, the weighted number of objects is defined as:
\begin{equation}
N_w=\sum_{i}^{N} w(G_{\rm BP}-G_{\rm RP},\, M_G,\,d)_i\times P_{i},
\end{equation}
where $N$ is the number of objects in that interval, $w$ the weighted function and $P_{i}$ the probability aforementioned. Error bars were estimated taking into account Poissonian error and the corresponding object weight, that is $\sigma_f=\sqrt{f\times (1-f)/N_W}$.

In Figure \ref{f:raw} we displayed our final spectral distribution (black solid circles and black line). To evaluate the extent of the completeness correction introduced in our sample, we included the ratio distribution of the entire classified white dwarf population (raw sample consisting of $65\,310$ objects) when no selection function is applied at all (blue open triangles). For comparative purposes, we also plotted the spectral distribution for the sample of 100\,pc \citep[red circles;][]{JE+2023}. 
The analysis of the different spectral distributions revealed that the effects due to completeness correction are of minor order. The general trend is practically coincident, and only small discrepancies appear for the coolest bin, where error bars underestimate the fact that the classification is less certain. In any case, we can conclude that our weighted spectral distribution provides a robust estimate of the ratio of DA versus non-DA white dwarfs.

\begin{figure*}[ht]
\centering

      {\includegraphics[width=0.76\textwidth, clip=true,trim=0 20 20 10]{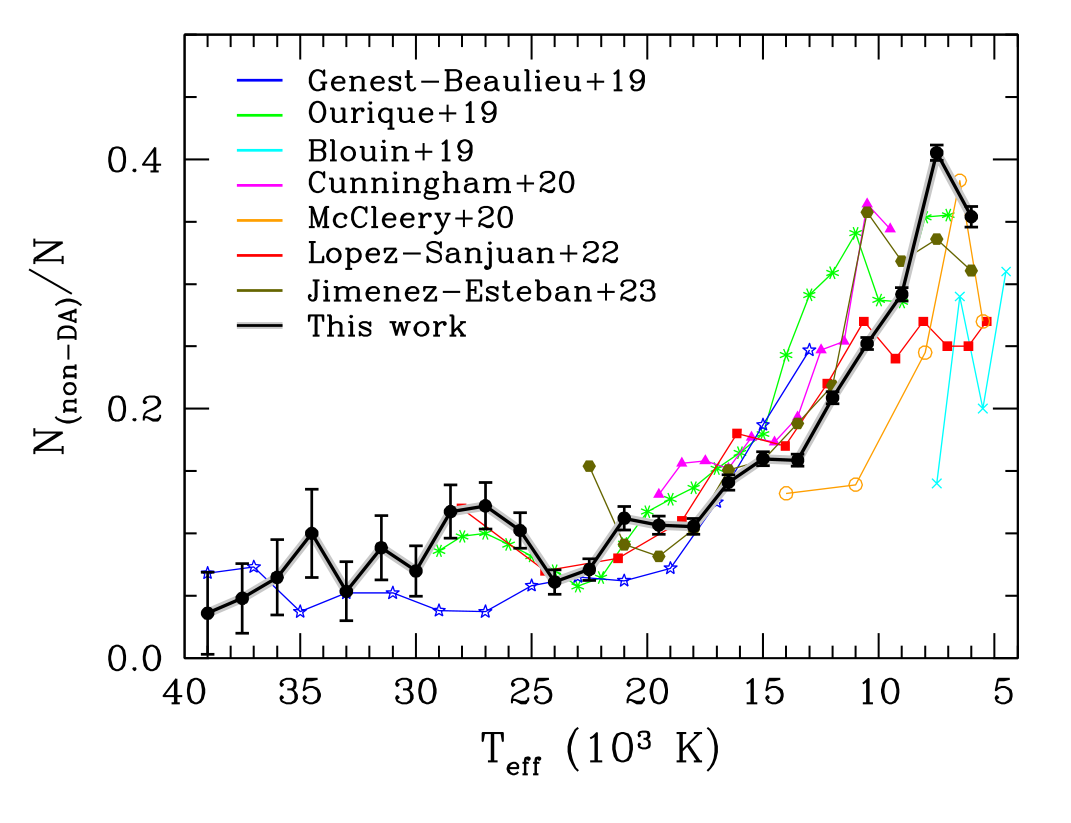}}

\caption{Spectral distribution for our final white dwarf sample (black points and lines). For comparative purposes we also show the spectral distributions from other works found in literature.}
\label{f:ratios}
\end{figure*}

In Figure \ref{f:ratios}, we show a comparison of the spectral distribution found in this work (black solid circles) with some of the most recent ratio distributions found in the literature. The first remarkable characteristic of the spectral distribution found in this work is the wider range of effective temperatures covered by it, thus constituting a solid estimate of the spectral evolution of white dwarfs. While a detailed analysis of the implications for spectral evolution is outside the scope of this work, in what follows, we make a brief analysis of the most relevant points found:
\begin{itemize}
\item[-] At the hottest end, i.e., $T_{\rm eff}\approx 35\,000-40\,000\,$K, the lowest ratio of non-DAs was found. As previously reported, there is not a complete absence of non-DAs in this region, but an average $\sim 5\%$ is indicative of the presence of the so called {\sl DB-gap} \citep[e.g.][and references therein]{Bergeron_2011,Koester2015}.
\item[-] A statistical significant deficit of non-DAs was also found for effective temperatures in the range $22\,000\lppr T_{\rm eff}\lppr 25\,000\,$K. It is also in perfect agreement with the ratios found by \citet{Ourique2018} and \citet{Lopez-Sanjuan2022}. 
\item[-] For temperatures cooler than $\sim18\,000$\,K, coinciding with the onset of convection mixing in DAs \citep[e.g.][]{Cunningham2020}, a marked increase  in the ratio of non-DAs was found, leading from $\sim10\%$ at $18\,000\,$K up to $\sim 40\%$ at $8\,000\,$K and in agreement with most of the spectral distributions found in the literature.
\item[-]  Our spectral distribution presented a peak around $T_{\rm eff}\approx7\,000\,$K with a ratio of non-DA objects of $f\approx 0.41$. This maximum is found at lower temperatures than that of our previous work \citep{JE+2023} and others retrieved in literature such as \citet{Ourique2020}, probably as a consequence of an unweighted distribution in these cases. However, the weighted distribution presented here is in agreement with the 40\,pc spectroscopic complete sample analyzed in \citet{McCleery2020}.
\item[-] A statistically significant decrease in the ratio of non-DAs is found at the coolest bin, $T_{\rm eff}\approx6\,000\,$K, dropping to $f\approx 0.35$. This behaviour was also reported in our previous work \citep{JE+2023} and is also in agreement with  \citet{Blouin2019} and \citet{McCleery2020}, although no known physical mechanism can be associated to it \citep{Blouin2019}.
\end{itemize}

\section{Conclusions}

Following the methodology presented in \citet{JE+2023}, we have expanded our white dwarf study sample up to 500 pc. A total of $65\,310$ white dwarfs have been classified as DAs and non-DAs based on their {\it Gaia} spectra, with an accuracy of 94\%. This has allowed us to construct a statistically significant and precise distribution of the DA versus non-DA ratio as a function of effective temperature. Nearly $34\,000$ white dwarfs have contributed to the final selected sample, making it the largest sample to date in terms of the number of objects and the range of effective temperatures analyzed, from $5\,500\,$K to $40\,000\,$K. 

The comparative analysis of our distribution with others found in the literature reveals statistically significant features such as: the deficit of DBs within the effective temperature range of approximately $35\,000-40\,000\,$K and between $22\,000--25\,000\,$K, along with a gradual rise starting from $18\,000\,$K up to around $7\,000\,$K, where the proportion of non-DA white dwarfs peaks at 41\%, followed by a decline for lower temperatures.

Finally, we can state that selection effects have been taken into account in the construction of the final sample. This fact, along with the high number of objects per interval in the sample, ensures that our spectral distribution can be considered a robust and precise element in the analysis of the spectral evolution of white dwarfs.

\begin{acknowledgements}
We acknowledge support from MINECO under the PID2020-117252GB-I00 grant and by the AGAUR/Generalitat de Catalunya grant SGR-386/2021. P.C. acknowledges financial support from the Government of Comunidad Autónoma de Madrid (Spain) via postdoctoral grant ‘Atracción de Talento Investigador’ 2019-T2/TIC-14760. R.M.O. is funded by INTA through grant PRE-OBSERVATORIO.  MC acknowledges
grant RYC2021-032721-I, funded by MCIN/AEI/10.13039/501100011033 and by the European Union NextGenerationEU/PRTR. RR acknowledges support from Grant RYC2021-030837-I funded by MCIN/AEI/ 10.13039/501100011033 and by “European Union NextGenerationEU/PRTR”. F.J.E. acknowledges support from ESA through the Faculty of the European Space Astronomy Centre (ESAC) - Funding reference 4000139151/22/ES/CM. This work has made use of data from the European Space Agency (ESA) mission {\it Gaia} (\url{https://www.cosmos.esa.int/gaia}), processed by the {\it Gaia} Data Processing and Analysis Consortium (DPAC, \url{https://www.cosmos.esa.int/web/gaia/dpac/consortium}). Funding for the DPAC has been provided by national institutions, in particular the institutions participating in the {\it Gaia} Multilateral Agreement. This work has made use of the Python package {\it GaiaXPy}, developed and maintained by members of the {\it Gaia} Data Processing and Analysis Consortium (DPAC) and in particular, Coordination Unit 5 (CU5), and the Data Processing Centre located at the Institute of Astronomy, Cambridge, UK (DPCI). This publication makes use of VOSA, developed under the Spanish Virtual Observatory (https://svo.cab.inta-csic.es) project funded by MCIN/AEI/10.13039/501100011033/ through grant PID2020-112949GB-I00. We extensively made used of Topcat \citep{Taylor05}. This research has made use of the VizieR catalogue access tool, CDS, Strasbourg, France. We acknowledge use of the ADS bibliographic services. 
\end{acknowledgements}


\bibliographystyle{aa}
\bibliography{wdev}


\end{document}